\documentclass[letterpaper, 12pt, onecolumn,final,conference]{ieeeconf}
	\IEEEoverridecommandlockouts                                                                                
	\usepackage{lipsum} 
	\usepackage{cite}
	\usepackage{graphicx}
	\usepackage{amsfonts,amssymb,amstext,verbatim,amsmath,graphicx,color,enumerate}
	\usepackage{bbold}
	
	\usepackage{tikz}
	\usepackage{pgfplots}
	\usepackage{float}
	 
	\usepackage{algorithm,algorithmic,xspace}
	
	\usepackage{epstopdf}
	
	\usepackage{bm}
	\usepackage{changes}

	\usetikzlibrary{arrows.meta}
	\usetikzlibrary{positioning, calc}
	\tikzset{
	 bus/.style={draw, circle, minimum size=2em,inner sep=0pt},
	 busg/.style={draw, circle, minimum size=2em,inner sep=0pt,color=black}
	}
	\usetikzlibrary{positioning,backgrounds}
	
	\newtheorem{theorem}{Theorem}[section]
	\newtheorem{proposition}[theorem]{Proposition}

	\newtheorem{assumption}[theorem]{Assumption}

	\newcommand{\idxi}{i}
	\newcommand{\idxj}{j}
	\newcommand{\idxl}{\ell}
	\newcommand{\idxm}{m}
	\newcommand{\idxh}{h}
	\newcommand{\idxt}{t}
	
	\newcommand{\graph}{G}
	\newcommand{\nodes}{\{1,\dots,N\}}
	\newcommand{\edges}{E}
	
	\newcommand{\nNodes}{N}
	\newcommand{\nEdges}{n}
	\newcommand{\inter}{S}
	\newcommand{\comm}{{\texttt{cmm}}}

	\newcommand{\Prob}{\mathbb{P}}
	\newcommand{\state}{\mathcal{C}}
	\newcommand{\nState}{C}
	\newcommand{\pState}{X}
	\newcommand{\oState}{x}
	\newcommand{\vState}{c}
	
	\newcommand{\test}{\mathcal{R}}
	\newcommand{\nTest}{R}
	\newcommand{\pTest}{Y}
	\newcommand{\oTest}{y}
	\newcommand{\boTest}{\bm{y}}
	\newcommand{\bpTest}{\bm{Y}}
	\newcommand{\vTest}{r}
	
	\newcommand{\ro}{\gamma}
	\newcommand{\bro}{\bm{\gamma}}
	\newcommand{\Ro}{\Gamma}
	\newcommand{\tensor}{\theta}
	\newcommand{\Tensor}{\Theta}
	\newcommand{\btensor}{\bm{\theta}}
	
	\newcommand{\IO}{\leftrightarrow}
	\newcommand{\ItoO}{\rightarrow}
	\newcommand{\OtoI}{\leftarrow}
	
	\newcommand{\LF}{L}

	\newcommand\oprocendsymbol{\hbox{$\square$}}
	\newcommand\oprocend{\relax\ifmmode\else\unskip\hfill\fi\oprocendsymbol}

	\definechangesauthor[color=red]{FS}
	\definechangesauthor[color=blue]{GN}
	\definechangesauthor[color=magenta]{AC}

	\def \final_ver{1}
	
	\newcommand{\RM}[2]{
		\ifnum \final_ver = 1
			\vphantom{#2}
		\else
			\deleted[id=#1]{#2}
		\fi
	}
	
	\newcommand{\AD}[2]{
		\ifnum \final_ver = 1
			{#2}
		\else
			\added[id=#1]{#2}
		\fi
	}

	\def \jphml/{JPH-ML}
	\def \djphml/{DJPH-ML}
	\def \djphfr/{DJPH-FR}
	
	\def \JPHNR/{JPH-NR}
	\def \JPHFR/{JPH-FR}

	\DeclareMathOperator*{\argmax}{argmax}
	
	\allowdisplaybreaks
	
	\title{Distributed Learning from Interactions in Social Networks}
	
	\author{Francesco Sasso, Angelo Coluccia, {\it Senior Member, IEEE}\\ and Giuseppe
	Notarstefano, {\it Member, IEEE} 
		\thanks{Francesco Sasso, Angelo Coluccia and Giuseppe Notarstefano are with the Department of
	Engineering, Universit\`a del Salento, via Monteroni, 73100, Lecce, Italy, \{name.lastname\}@unisalento.it.}
	\thanks{This result is part of a project that has received funding from the European Research Council
	(ERC) under the European Unions Horizon 2020 research and innovation
	programme (grant agreement No 638992 - OPT4SMART).
	}
	}

\begin{document}
	\bstctlcite{IEEEexample:BSTcontrol}
	
	\maketitle
	
	\begin{abstract}
	We consider a network scenario in which agents can evaluate each
	other according to a score graph that models some interactions. The goal is to design a distributed protocol, run by the agents,
	that allows them to learn their unknown state among a finite set of possible
	values. We propose a Bayesian framework in which scores and states are
	associated to probabilistic events with unknown parameters and hyperparameters,
	respectively. We show that each agent can learn its state by means of a local
	Bayesian classifier and a (centralized) Maximum-Likelihood (ML) estimator of
	parameter-hyperparameter that combines plain ML and Empirical Bayes
	approaches. By using tools from graphical models, which allow us to gain insight
	on conditional dependencies of scores and states, we provide a relaxed
	probabilistic model that ultimately leads to a parameter-hyperparameter
	estimator amenable to distributed computation. To highlight the
	appropriateness of the proposed relaxation, we demonstrate the distributed
	estimators on a social interaction set-up for user profiling.
	\end{abstract}

	\section{Introduction}
	
	A common feature of online social networks (OSNs) is the possibility of
	individuals to continuously interact among themselves, by sharing contents and
	expressing opinions or ratings on different topics \cite{Amelkin2017,
	  Tempo2017}.
	We address such a context by considering a  network scenario in which nodes can mutually
	rate, i.e., can give/receive a score to/from other ``neighboring'' nodes, and
	aim at learning their own (or their neighbors') state. The state may indicate a
	social orientation, influencing level, or the belonging to a thematic community.
	Due to the large-scale nature of OSNs, centralized solutions exhibit limitations both in terms of computation burden and
	privacy preservation, hence distributed solutions are needed.

	In recent years, a great interest has been devoted to distributed
	schemes in which nodes aim at estimating a common parameter, e.g., by means of
	Maximum Likelihood (ML) approaches,
	\cite{barbarossa2007decentralized,schizas2008consensus,chiuso2011gossip}, or
	performing simultaneous estimation and classification, \cite{chiuso2011gossip,fagnani2014distributed}.
	In \cite{coluccia2013distributed,coluccia2014hierarchical,coluccia2016bayesian} a more general Bayesian framework is considered,
	in which nodes estimate local parameters, rather than reaching consensus on a
	common one. In particular, an Empirical Bayes approach is proposed in which the parameters of the prior distribution,
	called \emph{hyperparameters}, are estimated through a distributed algorithm. The
	estimated hyperparameters are then combined with local measurements to obtain
	the Minimum Mean Square Error (MMSE) estimator of the local parameters.
	In the recent literature on distributed
	social learning, agents aim at estimating a \emph{common}
	unobservable state from noisy observations through non-Bayesian schemes in which each agent processes its
	own and its neighbors' beliefs
	\cite{jadbabaie2012non,shahrampour2013exponentially,lalitha2014social,molavi2016foundations,nedic2017fast},
	see also \cite{nedic2016tutorial} for a tutorial.
	A different batch of references investigates interpersonal influences in groups of individuals and
	the emerging of asymptotic opinions,
	\cite{mirtabatabaei2012opinion,mirtabatabaei2014reflected,friedkin2016network},
	see \cite{frasca2015distributed,Tempo2017} for a tutorial on opinion formation in social
	networks.
	The problem of self-rating in a social environment is discussed in
	\cite{li2016self}, where agents can perform a predefined task, but with
	different abilities.%

	In the present paper, we set up a learning problem in a network context in which
	each node needs to classify its own local state based on observations coming
	from the interaction with other nodes. Interactions among nodes are expressed by
	evaluations that a node performs on other ones, modeled through a weighted
	digraph that we will be referred to as \emph{score graph}. This general scenario
	captures a wide variety of contexts arising from social relationships, where
	nodes have only a partial knowledge of the world. Specifically, in
	Section~\ref{sec:setup} we devise a Bayesian probabilistic framework wherein,
	however, both the parameters of the observation model and the hyperparameters of
	the prior distribution are allowed to be unknown.
	In order to solve this interaction-based learning problem, we propose in
	Section~\ref{sec:algorithm} a learning approach combining a local Bayesian
	classifier with a joint parameter-hyperparameter Maximum Likelihood estimation
	approach.
	Since the ML estimator is computationally intractable even for moderately small
	networks, we resort to the conceptual tool of graphical models to identify a
	relaxation of the probabilistic model that leads to a distributed estimator.
	In Section~\ref{sub:community_discovery} we validate the performance of the
	proposed distributed estimator via Monte Carlo simulations.

	\section{Bayesian framework for interaction-based learning}\label{sec:setup}
	In this section, we set up the interaction-based learning problem in which
	agents of a network interact with each others according to a score graph. To
	learn its own state each node can use observations
	associated to incoming or outcoming edges. We propose a Bayesian probabilistic
	model with unknown parameters, which need to be estimated to solve
	the learning problem.
	
	\subsection{Interaction network model}
	We consider a \emph{network of agents} able to perform evaluations of other
	agents. The result of each evaluation is a score given by the evaluating agent
	on the evaluated one. Such an interaction is described by a \emph{score
	  graph}. Formally, we let $\nodes$ be the set of agent identifiers and
	$\graph_\inter = (\nodes,\edges_\inter)$ a digraph such that
	$(\idxi,\idxj)\in \edges_\inter$ if agent $\idxi$ evaluates agent $\idxj$.  We
	denote by $\nEdges$ the total number of edges in the graph, and assume that each
	node has at least one incoming edge in the score graph, that is, there is at
	least one agent evaluating it.
	
\begin{figure}
	\centering
	\includegraphics[scale=0.8]{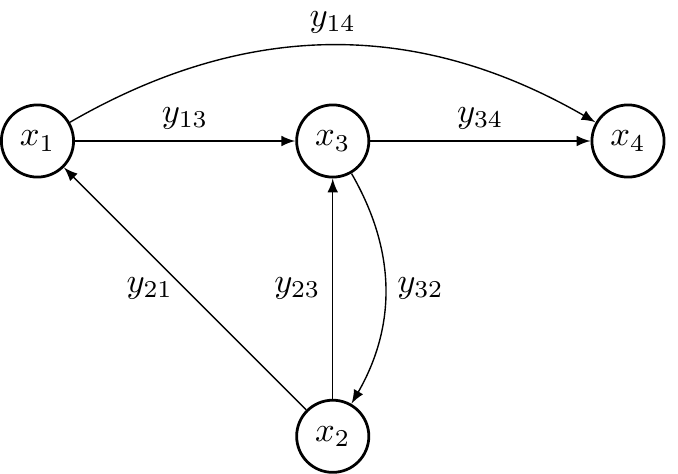}
	\caption{Example of a score graph $\graph_\inter$.}
	\label{fig:score-graph-example}% 
	\end{figure}

	Let $\state$ and $\test$ be the set of possible state and score values,
	respectively. Being finite sets, we can assume
	$\state = \{\vState_1,\dots,\vState_\nState\}$ and
	$\test = \{\vTest_1,\dots,\vTest_\nTest\}$, where $\nState$ and $\nTest$ are the
	cardinality of the two sets, respectively. Consistently, in the network we
	consider the following quantities:
	\begin{itemize}
	\item $\oState_\idxi \in \state$, unobservable \emph{state} (or community) of agent $\idxi$;
	\item $\oTest_{\idxi\idxj} \in \test$, \emph{score} (or evaluation result) of the evaluation performed by
	  agent $\idxi$ on agent $\idxj$.
	\end{itemize}
	An example of score graph with associated state and score values is shown in
	Fig.~\ref{fig:score-graph-example}.
	
	Besides the evaluation capability, the agents have also \emph{communication} and
	\emph{computation} functionalities. Agents communicate according to a
	time-dependent directed \emph{communication graph}
	$\idxt\mapsto\graph_\comm(\idxt) = (\nodes, \edges_\comm(\idxt))$, where the
	edge set $\edges_\comm(\idxt)$ describes the communication among agents:
	$(\idxi,\idxj)\in\edges_\comm(\idxt)$ if agent $\idxi$ communicates to $\idxj$
	at time $\idxt\in\mathbb{Z}_{\geq 0}$. We introduce the notation
	$\nNodes_{\comm,\idxi}^{I}(\idxt)$ and $\nNodes_{\comm,\idxi}^{O}(\idxt)$ for the
	in- and out-neighborhoods of node $\idxi$ at time $\idxt$ in the communication
	graph. We will require these neighborhoods to include the node $\idxi$ itself;
	formally, we have
	\begin{align*}
	\nNodes_{\comm,\idxi}^{I}(\idxt) &= \{\idxj : (\idxj,\idxi)\in\edges_\comm(\idxt)\}\cup\{\idxi\},\\
	\nNodes_{\comm,\idxi}^{O}(\idxt) &= \{\idxj : (\idxi,\idxj)\in\edges_\comm(\idxt)\}\cup\{\idxi\}
	\end{align*}
	
	For the communication graph we assume the following:
	\begin{assumption}\label{ass:connectivity}
	\! There exists an integer $Q\ge1$ such that the graph
	  $\bigcup_{\tau=\idxt Q}^{(\idxt+1)Q-1} \!\!\graph_\comm(\tau)\!$ is strongly connected $\forall \, t\!\ge\!0$.
	\end{assumption}
	
	We point out that in general the (time-dependent) communication graph, modeling
	the distributed computation, is not necessarily related to the (fixed) score
	graph. We just assume that when the distributed algorithm starts each node $i$
	knows the scores received by in-neighbors in the score graph.

	\subsection{Bayesian probabilistic model}
\begin{figure}
	\centering
	\includegraphics[scale=0.8]{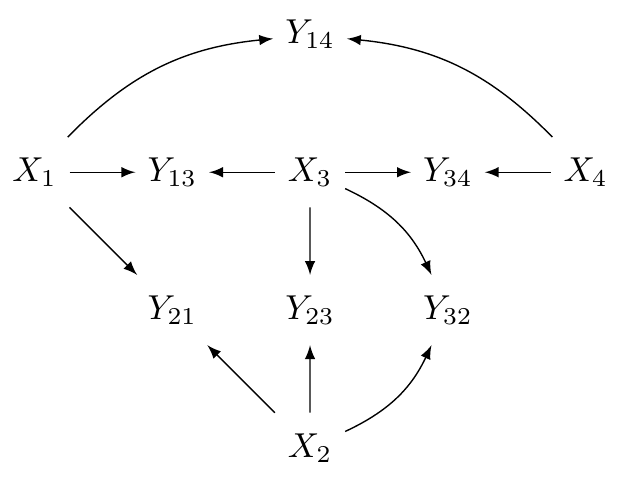}
	\caption{The score Bayesian network related to the score graph in Fig. \ref{fig:score-graph-example}.}
	\label{fig:graphical-model-example}%
	\end{figure}

	We consider the score $\oTest_{\idxi\idxj}, (\idxi,\idxj)\in\edges_\inter$, as
	the (observed) realization of a random variable denoted by
	$\pTest_{\idxi\idxj}$; likewise, each state value
	$\oState_{\idxi}, \idxi\in\nodes$, is the (unobserved) realization of a random
	variable $\pState_\idxi$.  To highlight the conditional dependencies
	among the random variables involved in the score graph, we resort to the tool of
	graphical models and in particular of Bayesian networks
	\cite{koller2009probabilistic}. Specifically, we introduce the \emph{Score
	  Bayesian Network} with $\nNodes+\nEdges$ nodes $\pState_\idxi$,
	$\idxi=1,\dots,N$, and $\pTest_{\idxi\idxj}$, $(\idxi,\idxj)\in\edges_\inter$ and
	$2\nEdges$ (conditional dependency) arrows defined as follows. For each
	$(\idxi,\idxj)\in\edges_\inter$, we have
	$\pState_\idxi\rightarrow\pTest_{\idxi\idxj}\leftarrow\pState_\idxj$ indicating
	that $\pTest_{\idxi\idxj}$ conditionally depends on $\pState_\idxi$ and
	$\pState_\idxj$.  In Fig.~\ref{fig:graphical-model-example} we represent the
	Score Bayesian Network related to the score graph in
	Fig.~\ref{fig:score-graph-example}.
	
	Denoting by $\bpTest_{\edges_\inter}$ the vector of all the random variables $\pTest_{\idxi\idxj},(\idxi,\idxj)\in\edges_\inter$, the joint distribution
	factorizes as
	\[
	\Prob(\bpTest_{\edges_\inter},\pState_1,\dots,\pState_\nNodes)\! = \!\Big(\prod_{(\idxi,\idxj)\in\edges_\inter}\!\!\!\!\Prob(\pTest_{\idxi\idxj}|\pState_\idxi,\pState_\idxj)\Big)\Big(\prod_{\idxi=1}^\nNodes\Prob(\pState_\idxi)\Big).
	\]
	We assume $\pTest_{\idxi\idxj}$, $(\idxi,\idxj)\in\edges_\inter$ are ruled by a
	conditional probability distribution
	$\Prob(\pTest_{\idxi\idxj} | \pState_\idxi,\pState_\idxj; \btensor)$, depending
	on a \emph{parameter} vector $\btensor$ whose components take values in a given
	set $\Tensor$. For notational purposes, we define the tensor
	\begin{equation}\label{eq:p_hlm}
	  p_{\idxh|\idxl,\idxm}(\btensor) := \Prob(\pTest_{\idxi\idxj} =
	  \vTest_\idxh | \pState_\idxi = \vState_\idxl,\pState_\idxj = \vState_\idxm;
	  \btensor),
	\end{equation}
	where $\vTest_\idxh\in\test$ and $\vState_\idxl, \vState_\idxm \in \state$.
	From the definition of probability distribution, we have  the constraint $\btensor\in\mathcal{S}_{\Tensor}$ with
	\begin{align*}
	\mathcal{S}_{\Tensor} := \Big\{\btensor\in\Tensor : p_{\idxh |
	  \idxl,\idxm}(\btensor)\in[0,1], \sum_{\idxh = 1}^{\nTest} p_{\idxh
	  | \idxl,\idxm}(\btensor) = 1\Big\}.
	\end{align*}
	
	We model $\pState_\idxi$, $\idxi=1,\dots,N$, as identically distributed random
	variables ruled by a probability distribution $\Prob(\pState_\idxi ; \bro)$,
	depending on a \emph{hyperparameter} vector $\bro$ whose components take
	values in a given set $\Ro$.  Again, we introduce the notation
	\begin{equation}\label{eq:p_l}
	p_\idxl(\bro) := \Prob(\pState_\idxi = \vState_\idxl; \bro).
	\end{equation}
	and, analogously to $\btensor$, we have the constraint
	$\bro \in \mathcal{S}_{\Ro}$ with
	\[
	\mathcal{S}_{\Ro} := \Big\{\bro\in\Ro : p_\idxl(\bro)
	\in [0,1], \sum_{\idxl = 1}^{\nState} p_\idxl(\bro) = 1 \Big\}.
	\]
	We assume that $p_{\idxh | \idxl,\idxm}$ and $p_\idxl$ are continuous functions, and
	that each node knows $p_{\idxh | \idxl,\idxm}, p_\idxl$ and the scores received
	from its in-neighbors and given to its out-neighbors in $\graph_\inter$.
	
	An example is discussed in the next subsection, while the problem of jointly
	estimating the \emph{parameter-hyperparameter} $(\btensor,\bro)$ will be
	addressed in Section~\ref{sec:algorithm}; the latter will be then a building
	block of the (distributed) learning scheme.

	\subsection{Example: social ranking scenario}\label{subsec:example_scenarios}
	
	A relevant scenario is user profiling in OSNs. In social relationships, in fact,
	people naturally tend to aggregate into groups based on some affinity; this is
	found also in OSN contexts.  For instance, consider a thread on a dedicated
	subject, wherein each member can express her/his preferences by assigning to
	other members/colleagues' posts a score from $1$ to $\nTest$ indicating an
	increasing level of appreciation for that post.
	To model the distribution of scores, we propose the following  variant of the
	so-called Mallow's $\phi$-model \cite{mallows1957nonnull}:
	\begin{equation}
	p_{\idxh|\idxl,\idxm}(\tensor) = \frac{1}{\psi_{\idxl,\idxm}(\tensor)} e^{-\big(\frac{(\vTest_\nTest -
		  \vTest_\idxh) / \vTest_\nTest - d(\vState_\idxl,\vState_\idxm) /
		  \vState_\nState}{\tensor}\big)^2},
	\label{eq:mallow}
	\end{equation}
	where $\vTest_\idxh = \idxh$ ($\idxh = 1,\dots,\nTest$), $\vState_\idxl = \idxl$
	($\idxl = 1,\dots,\nState$), $\tensor\in\mathbb{R}_{>0}$ is a dispersion parameter,
	$\psi_{\idxl,\idxm}(\tensor)$ is a normalizing constant, and $d$ is a
	semi-distance, i.e., $d \ge 0$ and $d(\vState_\idxl,\vState_\idxm) = 0$ if and only if
	$\vState_\idxl=\vState_\idxm$.
	Informally, the ``farther'' a given community $\vState_\idxl$ is from another
	community $\vState_\idxm$, the higher will be the distance
	$d(\vState_\idxl,\vState_\idxm)$, and thus the lower the score.
	
	In many cases the resulting subgroups
	reflect some hierarchy in the population: basic examples are forums or
	working teams. Thus, we consider a scenario in which each person belongs to a community
	reflecting some degree of expertise about a given topic or field.  In
	particular, we have $\nState$ ordered communities, with $\ell$th community given
	by $\vState_\idxl = \idxl$. That is, for example, a person in the community
	$\vState_1$ is a \emph{newbie}, while a person in $\vState_\nState$ is a
	\emph{master}.
	Since climbing in the hierarchy can be regarded as the result of several ``promotion''
	events, a possible probabilistic model for the communities is a binomial
	distribution $\mathcal{B}(\nState - 1,\ro)$, where $\ro\in[0,1]$ represents the
	probability of being promoted, i.e.,
	\begin{align*}
	p_\idxl(\ro) &= \binom{\nState - 1}{\vState_\idxl - 1}\ro^{\vState_\idxl - 1}(1
				   - \ro)^{\nState - 1 - (\vState_\idxl - 1)}.
	\end{align*}
	We will refer to this set-up as \emph{social-ranking model}.

	\section{Interaction-based distributed learning}\label{sec:algorithm}
	In this section we describe the proposed distributed learning scheme.
	Without loss of generality, we focus on a set-up in which a node wants to
	self-classify. The same scheme also applies to a scenario in which a node wants
	to classify its neighbors, provided it knows their given and received scores.
	
	The section is structured as follows.
	First, we derive a local Bayesian classifier provided that an estimation of
	parameter-hyperparameter $(\btensor,\bro)$ is available.
	Then, based on a combination of plain ML and Empirical Bayes estimation
	approaches, we derive a joint parameter-hyperparameter estimator.
	Finally, we propose a suitable relaxation of the Score Bayesian Network which
	leads to a distributed estimator, based on proper distributed
	optimization algorithms.

	\subsection{Bayesian classifiers (given parameter-hyperparameter)}
	Each node can self-classify (i.e., learn its own state) if an estimate
	$(\bm{\hat{\tensor}}, \bm{\hat{\ro}})$ of
	parameter-hyperparameter $(\btensor, \bro)$ is
	available. Before discussing in details how this estimate can be obtained in a distributed way, we develop a \emph{decentralized} MAP self-classifier
	which uses only single-hop information, i.e., the scores it gives to and receives from neighbors.

\begin{figure}
	\centering
		\includegraphics[scale=0.7]{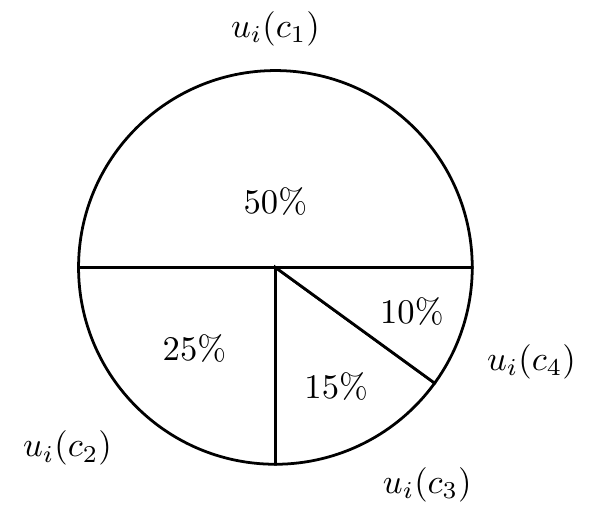}
	\caption{Example of outcome of the soft classifier of an agent $i$, for $\nState = 4$: $\bm{u}_\idxi = (0.5,0.25,0.15,0.1)$.}
	\label{fig:soft-classifier}
	\end{figure}

	Formally, let $\bm{\oTest}_{N_\idxi}$ be the vector of (observed) scores that
	agent $\idxi$ obtains by in-neighbors and provides to out-neighbors, i.e., the
	stack vector of $\oTest_{\idxj\idxi}$ with $(\idxj,\idxi)\in\edges_\inter$ and
	$\oTest_{\idxi\idxj}$ with $(\idxi,\idxj)\in\edges_\inter$. Consistently, let
	${\bm{\pTest}\!}_{N_\idxi}$ be the corresponding random vector. For each
	agent $\idxi=1,\dots,\nNodes$, we define
	\begin{equation*}
	u_\idxi(\vState_\idxl) := \Prob(\pState_\idxi = \vState_\idxl |
	{\bm{\pTest}\!}_{N_\idxi} = \bm{\oTest}_{N_\idxi}; \bm{\hat{\ro}},
	\bm{\hat{\tensor}}), \quad \idxl = 1,\dots,\nState.
	\end{equation*}
	The \emph{soft classifier} of $\idxi$ is the probability vector
	$\bm{u}_\idxi := (u_\idxi(\vState_1),\dots,u_\idxi(\vState_\nState))$
	(whose components are nonnegative and sum to $1$). In
	Fig.~\ref{fig:soft-classifier} we depict a pie-chart representation of an example vector
	$\bm{u}_\idxi$.
	
	From the soft classifier we can define the classical \emph{Maximum A-Posteriori probability (MAP) classifier} as the argument corresponding to the maximum
	component of $u_\idxi$, i.e.,
	\begin{equation*}
	\hat{\oState}_\idxi := \argmax_{\vState_\idxl\in\state} u_\idxi(\vState_\idxl).
	\end{equation*}
	
	The main result here is to show how to efficiently compute the MAP classifiers. First,
	we define
	\begin{align*}
	\nNodes_\idxi^{\IO}&:=\{\idxj:(\idxj,\idxi)\in\edges_\inter,(\idxi,\idxj)\in\edges_\inter\},\\
	\nNodes_\idxi^{\OtoI}&:=\{\idxj:(\idxj,\idxi)\in\edges_\inter,(\idxi,\idxj)\notin\edges_\inter\},\\
	\nNodes_\idxi^{\ItoO}&:=\{\idxj:(\idxi,\idxj)\in\edges_\inter, (\idxj,\idxi)\notin\edges_\inter\},
	\end{align*}
	and for each $h,k=1,\dots,R$ we introduce the quantities:
	\begin{align*}
	n_\idxi^{\IO}(h,k) &:= |\{j\in\nNodes_\idxi^{\IO}: y_{ij}=r_h, y_{ji}=r_k\}|,\\
	n_\idxi^{\OtoI}(h) &:= |\{j\in\nNodes_\idxi^{\OtoI}: y_{ji}=r_h\}|,\\
	n_\idxi^{\ItoO}(h) &:= |\{j\in\nNodes_\idxi^{\ItoO}: y_{ij}=r_h\}|.
	\end{align*}
	
	\begin{theorem}\label{thm:MAP_characterization}
	  Let $\idxi\in\nodes$ be an agent of the score graph. Then, the
	  components of the vector $u_\idxi$ are given by
	\begin{equation*}
		u_{\idxi}(\vState_\idxl) = \frac{v_{\idxi}(\vState_\idxl)}{Z_{\idxi}}
	\end{equation*}
	where $Z_{\idxi} = \sum_{\ell = 1}^C v_{\idxi}(\vState_\idxl)$ is a normalizing
	constant, and
	$v_{\idxi}(\vState_\idxl) =
	p_\idxl(\bm{\hat{\ro}})\pi_\idxi^{\IO}(c_\idxl)\pi_\idxi^\OtoI(c_\idxl)\pi_\idxi^\ItoO(c_\idxl)$
	with
	\begin{align*}
	\pi_\idxi^{\IO}(c_\idxl) &= \prod_{h,k=1}^C\Big(\sum_{m=1}^C p_{k|\idxm,\idxl}(\bm{\hat{\tensor}})p_{h|\idxl,\idxm}(\bm{\hat{\tensor}})p_{\idxm}(\bm{\hat{\ro}})\Big)^{n_\idxi^{\IO}(h,k)},\\
	\pi_\idxi^\OtoI(c_\idxl) &= \prod_{h = 1}^R\Big(\sum_{\idxm=1}^C p_{h|\idxm,\idxl}(\bm{\hat{\tensor}})p_\idxm(\bm{\hat{\ro}})\Big)^{n_\idxi^\OtoI(h)},\\
	\pi_\idxi^\ItoO(c_\idxl) &= \prod_{h = 1}^R\Big(\sum_{\idxm=1}^C p_{h|\idxl,\idxm}(\bm{\hat{\tensor}})p_\idxm(\bm{\hat{\ro}})\Big)^{n_\idxi^\ItoO(h)}.
	\end{align*}
	\oprocend
	\end{theorem}
	The proof is given in \cite{sasso2017interaction}.

	\subsection{Joint Parameter-Hyperparameter ML estimation (JPH-ML)}
	Classification requires that at each node an estimate
	$(\bm{\hat{\tensor}}, \bm{\hat{\ro}})$ of
	parameter-hyperparameter $(\btensor, \bro)$ is available.

	On this regard, a few remarks about $\btensor$ and $\bro$ are now in
	order. Depending on both the application and the network context, these
	parameters may be known, or (partially) unknown to the nodes. If both of them
	are known, we are in a pure Bayesian set-up in which, as just shown, each node
	can independently self-classify with no need of cooperation. The case of unknown
	$\btensor$ (and known $\bro$) falls into a Maximum-Likelihood framework, while
	the case of unknown $\bro$ (and known $\btensor$) can be addressed by an
	\emph{Empirical Bayes} approach.
	In this paper we consider a general scenario in which both of them
	can be unknown.
	Our goal is then to compute, in a distributed way, an estimate of
	\emph{parameter-hyperparameter} $(\btensor,\bro)$ and use it for the
	classification at each node.
	In the following we show how to compute it in a distributed way by following a mixed
	Empirical Bayes and Maximum Likelihood approach.
	The \emph{Joint Parameter-Hyperparameter Maximum Likelihood (\jphml/)
	  estimator} can be defined as
	\begin{equation}
	\label{eq:MLE}
	(\bm{\hat{\tensor}}_{\text{\tiny ML}}, \bm{\hat{\ro}}_{\text{\tiny ML}}) :=
	\argmax_{(\btensor, \bro)\in
	  \mathcal{S}_{\Tensor}\times\mathcal{S}_{\Ro}}
	\LF({\bm{\oTest}\!}_{\edges_\inter};\btensor, \bro)
	\end{equation}
	where ${\bm{\oTest}\!}_{\edges_\inter}$ is the vector of all
	scores $\oTest_{\idxj\idxi}, (\idxj,\idxi)\in\edges_\inter$, and
	\begin{equation}
	\label{eq:likelihood}
	\LF({\bm{\oTest}\!}_{\edges_\inter};\btensor, \bro) =
	\Prob({\bm{\pTest}\!}_{\edges_\inter} = {\bm{\oTest}\!}_{\edges_\inter} \, ; \, \btensor, \bro)
	\end{equation}
	is the \emph{likelihood function}.

	Notice that $\btensor$ is directly linked to the observables
	${\bm{\oTest}\!}_{\edges_\inter}$; the hyperparameter $\bro$ is instead related to the
	unobservable states. While one could readily obtain the likelihood
	function for the sole estimation of $\btensor$ from the distribution of scores,
	the presence of $\bro$ requires to marginalize over all unobservable state
	(random) variables.
	By  the law of total probability
	\begin{align}
	\begin{split}
	&\LF({\bm{\oTest}\!}_{\edges_\inter};\btensor, \bro) = \\
	&\hspace{5pt}\sum_{\idxl_1=1}^\nState\! \cdots\! \sum_{\idxl_\nNodes = 1}^\nState
		  \!\Prob({\bpTest\!}_{\edges_\inter}\!=\! {\boTest\!}_{\edges_\inter},
		   \pState_{1}\!=\!\vState_{\idxl_1},\dots,\pState_{\nNodes}\!=\!\vState_{\idxl_\nNodes}).
	\end{split}
	\label{eq:LF_deriv_1}
	\end{align}
	Indicating with $\nNodes^{I}_{\idxi}$ the set of in-neighbors of agent $\idxi$
	in the score graph (we are assuming that it is non-empty), the probability in
	\eqref{eq:LF_deriv_1} can be written as the product of
	the conditional probability of scores, i.e.,
	\begin{align*}
	  \Prob({\bpTest\!}_{\edges_\inter}= {\boTest\!}_{\edges_\inter}\,|\,
			  &\pState_{1}=\vState_{\idxl_1},\dots, \pState_{\nNodes}=\vState_{\idxl_\nNodes})
			  = \\
	  &\prod_{\idxi =
	  1}^\nNodes\prod_{\idxj\in\nNodes^{I}_{\idxi}} \Prob(\pTest_{\idxj\idxi} =
	  \oTest_{\idxj\idxi}| \pState_{\idxj} = \vState_{\idxl_\idxj},
	  \pState_{\idxi} = \vState_{\idxl_\idxi})
	\end{align*}
	multiplied by the prior probability of states, i.e.,
	\begin{align*}
	\Prob(\pState_{1}=\vState_{\idxl_1},\dots, \pState_{\nNodes}=\vState_{\idxl_\nNodes})
	  = \prod_{\idxi=1}^{\nNodes}\Prob(\pState_{\idxi}=\vState_{\idxl_\idxi}).
	\end{align*}
	Thus, the likelihood function turns out to be
	\begin{align*}
	  \LF({\bm{\oTest}\!}_{\edges_\inter};\btensor, \bro)
	\!=\! \sum_{\idxl_1=1}^\nState \!\cdots \!\sum_{\idxl_\nNodes =
	  1}^\nState\prod_{\idxi=1}^\nNodes
	p_{\idxl_\idxi}(\bro)\prod_{\idxj\in\nNodes^{I}_{\idxi}}p_{h_{ji}\,|\,
	  \idxl_\idxi,\idxl_\idxj }(\btensor)
	\end{align*}
	where $\idxh_{\idxj\idxi}$ is the index of the score element
	$\vTest_{\idxh_{\idxj\idxi}} \in \test=\{\vTest_1,\ldots,\vTest_\nTest\}$ associated to the score
	$\oTest_{\idxj\idxi}$, i.e., $\oTest_{\idxj\idxi}= \vTest_{\idxh_{\idxj\idxi}}$.

	\subsection{Distributed JPH Node-based Relaxed estimation (JPH-NR)}
\begin{figure}
		\centering
		\includegraphics{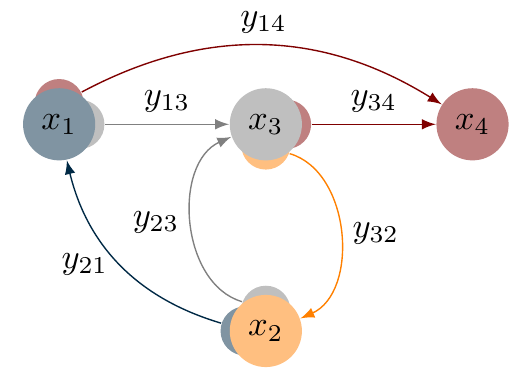}
		\caption{Node-based relaxation of the score graph in Fig. \ref{fig:score-graph-example}, with virtual nodes indicating the virtual states of each node.}
			\label{fig:score-graph-middle}
	\end{figure}

\begin{figure}
		\centering
		\includegraphics{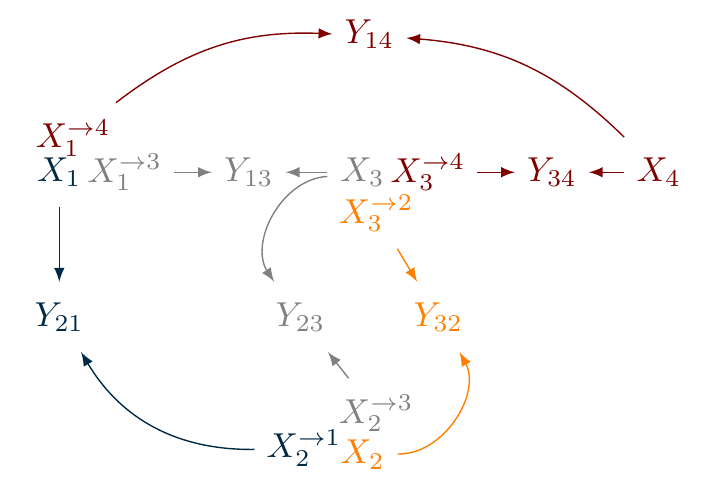}
		\caption{Node-based relaxation of the score Bayesian network of Fig. \ref{fig:score-graph-middle}.}
			\label{fig:graphical-model-middle}
	\end{figure}

	From the equations above it is apparent that the likelihood function couples the
	information at all nodes, so problem~\eqref{eq:MLE} is not amenable to
	distributed solution.  To make it distributable, we propose a relaxation
	approach.  To this aim we introduce, instead of
	$\LF({\bm{\oTest}\!}_{\edges_\inter};\btensor, \bro)$, a \emph{Node-based
	  Relaxed (NR) likelihood}
	$\LF_{NR}({\bm{\oTest}\!}_{\edges_\inter};\btensor, \bro)$.
	Let $\bm{\oTest}_{N_\idxi^I}$ be the vector of (observed) scores that agent
	$\idxi$ obtains by in-neighbors and ${\bm{\pTest}\!}_{N_\idxi^I}$ the
	corresponding random vector. Then,
	\begin{align}
	 \LF_{NR}({\bm{\oTest}\!}_{\edges_\inter};\btensor, \bro)
	:= \prod_{\idxi = 1}^\nNodes \Prob({\bm{\pTest}\!}_{N_\idxi^I} = {\bm{\oTest}\!}_{N_\idxi^I} \, ; \, \btensor,\bro).
	\end{align}
	This relaxation can be interpreted as follows. We imagine that each node has a
	virtual state, independent of its true state, every time it evaluates another
	node. Thus, in the Score Bayesian Network, besides the state variables
	$\pState_\idxi$, $\idxi = 1,\dots,N$, there will be additional variables
	$\pState_\idxi^{\rightarrow \idxj}$ for each $j$ with $(i,j)\in\edges_\inter$.
	To clarify this model,
	Figs.~\ref{fig:score-graph-middle}-\ref{fig:graphical-model-middle} depict the
	node-based relaxed graph and the corresponding graphical model for the same example
	given in Figs.~\ref{fig:score-graph-example}-\ref{fig:graphical-model-example}.
	

	Since ${\bm{\pTest}\!}_{N_\idxi^I}$, $\idxi=1,\dots,N$, are not independent, then
	clearly $\LF\neq\LF_{NR}$.
	However, as it will appear from the numerical performance assessment reported in
	the Section~\ref{sub:community_discovery}, this choice yields reasonably small
	estimation errors.
	
	Using this virtual independence between ${\bm{\pTest}\!}_{N_\idxi^I}$, with
	$\idxi=1,\dots,N$, we define the \emph{JPH-NR estimator} as
	\begin{equation}
	\label{eq:MLE-NR}
	(\bm{\hat{\tensor}}_{\text{\tiny NR}}, \bm{\hat{\ro}}_{\text{\tiny NR}}) :=
	\argmax_{(\btensor, \bro)\in
	  \mathcal{S}_{\Tensor}\times\mathcal{S}_{\Ro}}
	\LF_{NR}({\bm{\oTest}\!}_{\edges_\inter};\btensor, \bro).
	\end{equation}
	
	The next result characterizes the structure of \JPHNR/ \eqref{eq:MLE-NR}.
	\begin{proposition}
	\label{prop:loglike_NR}
	The \JPHNR/ estimator based on the node-based relaxation of the
	score Bayesian network is given by
	\begin{equation}
	\label{eq:NRopt}
	(\bm{\hat{\tensor}}_{\text{\tiny \textnormal{NR}}}, \bm{\hat{\ro}}_{\text{\tiny \textnormal{NR}}}) =
	\argmax_{(\btensor, \bro)\in
	  \mathcal{S}_{\Tensor}\times\mathcal{S}_{\Ro}}
	\sum_{\idxi=1}^{\nNodes} g(\btensor,\bro;\bm{\nEdges}_i)
	\end{equation}
	with $\bm{\nEdges}_i \!=\! [\nEdges_\idxi^{(1)} \cdots
	\nEdges_\idxi^{(\nTest)}]^\top$,  \!\! $\nEdges_\idxi^{(\idxh)} \!:=\! |\{\idxj\in
	N^I_\idxi: \! y_{\idxj\idxi}\! =\! r_h\}|$, \!\! and
	\begin{align}\label{eq:g}
	g(\btensor,\bro;\bm{n}_i) &=\nonumber\\
	\log\Big(\sum_{\idxl = 1}^C&  p_\idxl(\bro)\prod_{h =
	  1}^R \Big(\sum_{m=1}^C p_{h \,|\, m, \idxl}(\btensor) p_m(\bro)\Big)^{n_i^{(h)}}\Big).
	\end{align}
	\oprocend
	\end{proposition}
	The proof is given in \cite{sasso2017interaction}.

	Proposition~\ref{prop:loglike_NR} ensures that the \JPHNR/ estimator can be
	computed by solving an optimization problem that has a separable cost (i.e., the
	sum of $N$ local costs). Available distributed optimization algorithms for
	asynchronous networks can be adopted to this aim, e.g.
	\cite{carli2015analysis}, \cite{nedic2015distributed}, \cite{di2016next}.

	\section{Distributed learning for social ranking}\label{sub:community_discovery}
	
	\begin{figure}
		\centering
		\includegraphics[scale=0.7]{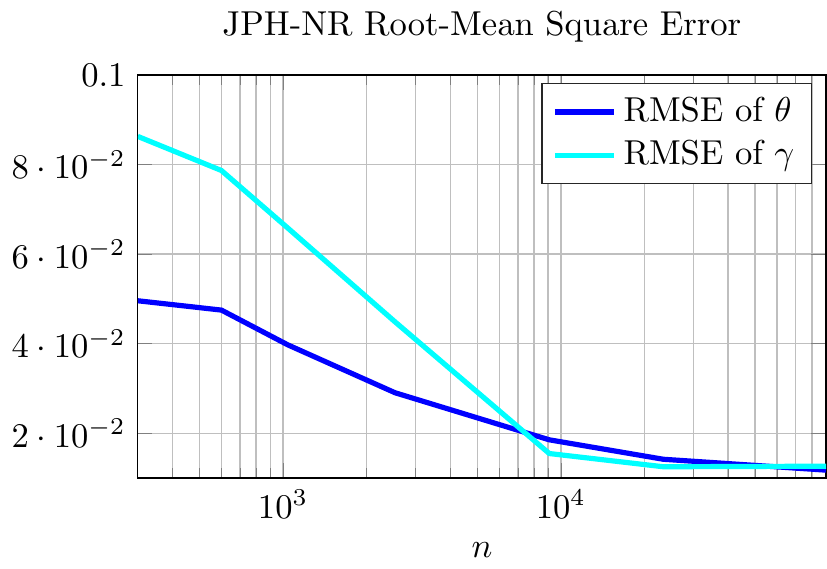}
		\caption{\JPHNR/ RMSE of the estimates of $(\tensor,\ro)$ as a function of the number of edges $\nEdges$.}
		\label{fig:ranking-100-theta}
	\end{figure}

	We report numerical results for the social ranking model described in
	Section~\ref{subsec:example_scenarios} with $\nState=6$, $\nTest=3$ and $\nNodes = 300$.
	We adopt in \eqref{eq:mallow} the semi-distance
	$ d(\vState_\idxl,\vState_\idxm) = |\vState_\idxl - \vState_\idxm| = |\idxl -
	\idxm|$. The true values of parameter-hyperparameter are $\tensor = \frac{1}{5}$ and $\ro = \frac{3}{10}$.

	Monte Carlo simulations have been run to test the performance of the \JPHNR/ estimator, with $1000$ trials for each point.
	 Fig.~\ref{fig:ranking-100-theta} reports
	the RMSE for the estimation of $(\tensor,\ro)$ as a function of the number of
	edges. It is worth noting that the estimation errors decrease as the number
	of edges increases, since more data are available.

	\begin{figure}
		\centering
		\includegraphics[scale=0.7]{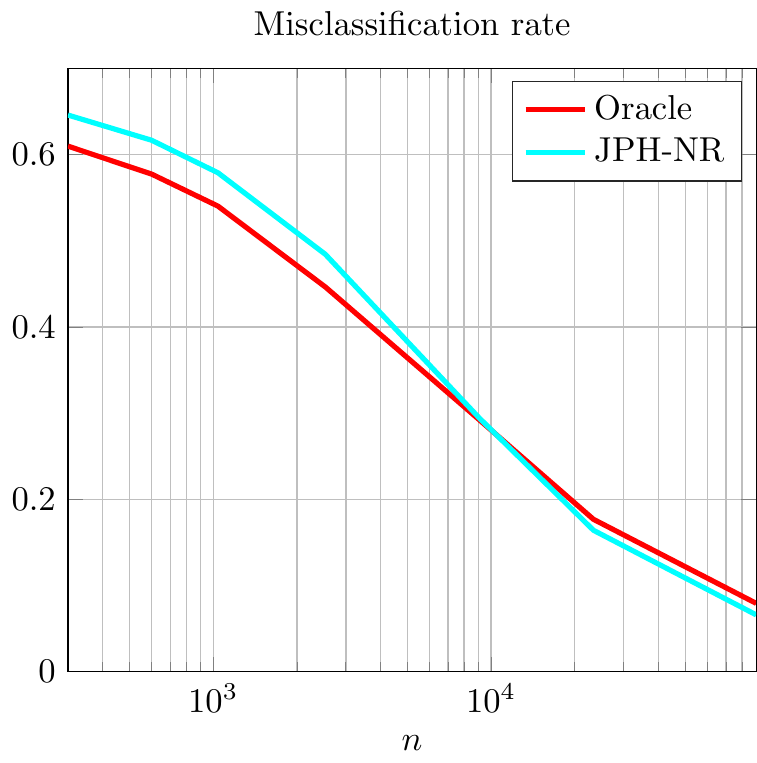}
		\caption{Misclassification rate as function of the number of edges $\nEdges$
		increasing from $\nNodes$ (cyclic graph) to $\nNodes^2-\nNodes$ (complete
		graph), with $\nNodes = 300$, $\ro = \frac{3}{10}$, $\tensor = \frac{1}{5}$, $C = 6$ and $R = 3$.}
		\label{fig:ranking-100}
	\end{figure}
	
	\begin{figure}
	\centering
		\includegraphics[scale=0.7]{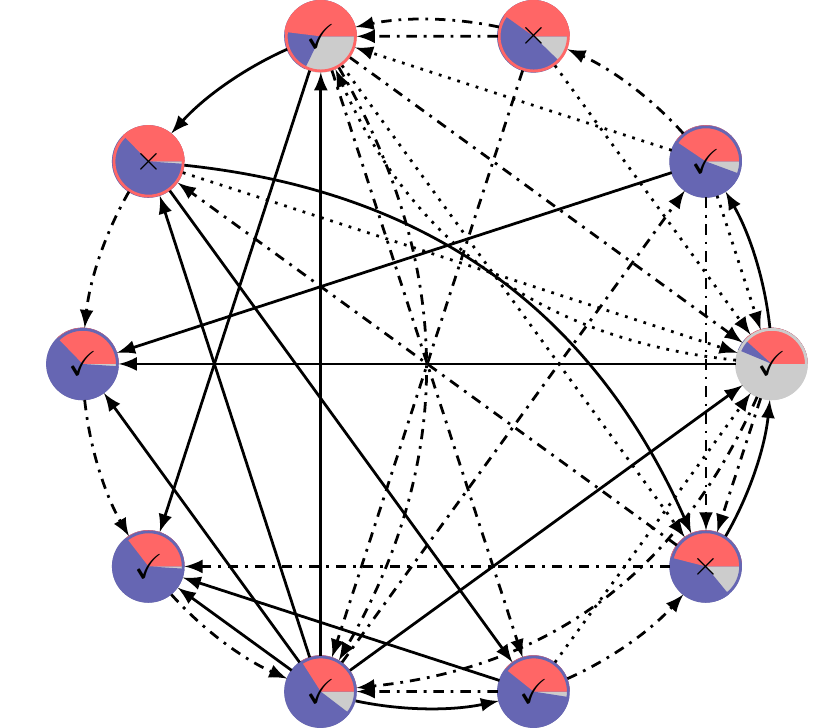}
	\caption{Soft classifier representation of a particular score graph.}
	\label{fig:ranking-soft}
	\end{figure}
	
	The impact of estimation errors on
	the learning performance is shown in Fig.~\ref{fig:ranking-100}: the curve
	clearly shows that the inferential
	relationship between scores and states is ``weaker'' hence more data are needed
	for a good learning. As a benchmark, the curve corresponding also to the ``oracle'' classifier that uses the true value of $\ro$ and $\tensor$ is reported. Remarkably, the proposed estimator is very close
	to the performance of the benchmark.

	Finally, we report an additional case to highlight the usefulness of the soft
	classifier. We considered a network of $\nNodes = 10$ agents, divided in $C = 3$ communities, in which the maximum score is $R = 3$. The related score graph
	$\graph_\inter$ is shown in Fig.  \ref{fig:ranking-soft}. We drew the states and
	scores in the given score graph according to the previous distributions, and
	then used the social ranking model to solve the learning problem as before, by
	means of the \JPHNR/ estimator.

	The contour of a node has a color which indicates the true state of the
	node. Inside the node we have represented the outcome of the soft
	classification, i.e., the output of the local self-classifier, as a
	pie-chart. The colors used are: red for state $c_1$, blue for state $c_2$, gray for
	state $c_3$. Moreover, each edge is depicted by a different pattern based on its
	evaluation result $r_h$: solid lines are related to scores equal to $3$, dash
	dot lines are related to scores equal to $2$, while dotted lines are related to
	scores equal to $1$. We assigned to each node a symbol $\checkmark$ or $\times$
	indicating if the MAP classifier correctly decided for the true state or not.
	
	Fig. \ref{fig:ranking-soft} shows a realization with three misclassification errors; remarkably, all of them
	correspond to a lower confidence level given by the soft classifier, which is an
	important indicator of the lack of enough information to reasonably trust the
	decision. It can be observed that the edge patterns concur to determine the
	decision. Indeed, the only gray-state node is correctly classified thanks to the
	predominant number of dotted edges insisting on it, and similarly for the
	blue-state nodes which mostly have solid incoming edges. When a mix of scores
	are available, clearly there is more uncertainty and the learning may fail, as
	for two of the red-state nodes.

	\section{Conclusion}\label{sec:conclusions}
	In this paper we have proposed a novel probabilistic framework for distributed
	learning, which is particularly relevant to emerging contexts such as
	cyber-physical systems and social networks. In the proposed set-up, nodes of a
	network want to learn their (unknown) state; differently from a classical
	set-up, the information does not come from (noisy) measurements of the state but
	rather from observations produced by the interaction with other nodes. For this
	problem we have proposed a hierarchical (Bayesian) framework in which the
	parameters of the interaction model as well as hyperparameters of the prior
	distributions may be unknown. Node classification is performed by means of a
	local Bayesian classifier that uses parameter-hyperparameter estimates, obtained
	by combining the plain ML with the Empirical Bayes estimation approaches in a
	joint scheme. The resulting estimator is very general but, unfortunately, not
	amenable to distributed computation. Therefore, by relying on the conceptual
	tool of graphical models, we have proposed an approximated ML estimator that
	exploits a proper relaxation of the conditional dependencies among the involved
	random variables. Remarkably, the approximated likelihood function leads to
	distributed estimation algorithms. To demonstrate the application of the
	proposed schemes, we have addressed an example scenario from user profiling in
	social networks, for which Monte Carlo simulations are reported. Results show
	that the proposed distributed learning scheme, although based on relaxation of
	the exact likelihood function, exhibits performance very close to the ideal
	classifier that has perfect knowledge of all parameters.

	\bibliographystyle{IEEEtran}
	\bibliography{distributed_classification}

	\end{document}